\documentclass[fleqn,twoside]{article}
\usepackage{espcrc2}
\usepackage{graphicx}
\usepackage{dcolumn}
\usepackage{bm}
\usepackage{epsfig}

\def\alt{\hbox{\hskip 0.4em\lower 0.7ex\hbox{$\sim$}%
\kern-0.8em\raise 0.45ex\hbox{$<$}\hskip 0.4em}}
\def\agt{\hbox{\hskip 0.4em\lower 0.7ex\hbox{$\sim$}%
\kern-0.8em\raise 0.45ex\hbox{$>$}\hskip 0.4em}}

\def\gev{GeV~c$^{-2}$}

\begin{document}

\title{ 
Expected Performance of CryoArray
}

\author{R.W.~Schnee\address[cwru]{Department of Physics, 
        Case Western Reserve University, Cleveland, OH 44106, USA},
        D.S.~Akerib\addressmark[cwru],
      R.J.~Gaitskell\address{Department of Physics, Brown University,   
		  Providence, RI 02912, USA} 
}

\begin{abstract}
WIMP-nucleon cross sections $\sigma \alt 10^{-9}$~pb
may be probed by ton-scale experiments with low thresholds and background 
rates $\sim 20$ events per year.
An array of cryogenic detectors (``CryoArray'') could perform 
well enough to reach this goal.  
Sufficient discrimination and background suppression of photons has 
already been demonstrated.  Reduction of neutron backgrounds may be 
achieved by siting the experiment deep enough.  Removal of the 
surface-electron backgrounds alone has not yet been demonstrated, 
but the reductions required even for this troublesome background 
are quite modest and appear achieveable.
\end{abstract}

\maketitle

\section{Introduction}

\begin{figure}
\begin{center}
\includegraphics [width=.403\textwidth]{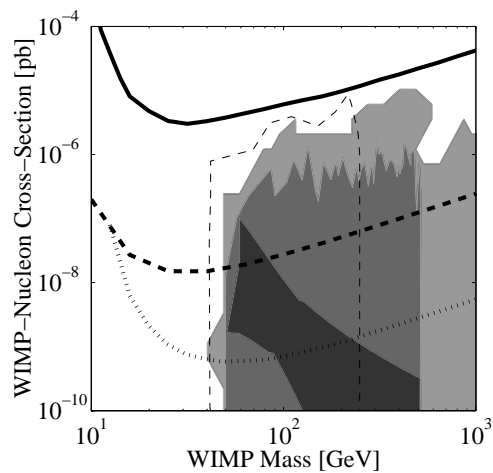}
\end{center}
\caption{Comparison of theoretical expectations and experimental
sensitivities for spin-independent WIMP-nucleon cross section vs WIMP mass. 
Curves indicate experimental limits for CDMS I~\cite{r19prlprd} (solid), 
and projected sensitivities for CDMS II at Soudan (dashed) and 
for CryoArray (dotted).
The region outlined in dashes~\cite{bottino}
and the lightest shaded region~\cite{baltz01} each shows the 
results from
calculations under an effective supersymmetry theory.
The medium-gray and darkest regions~\cite{ellis01} arise from more 
constrained frameworks.
These and other results and projections 
are available via an interactive web plotter~\protect\cite{dmplotter}.
 }
\label{limitplot}
\end{figure}

Direct detection of supersymmetric WIMP dark matter in the coming 
decade appears possible.
As shown in Fig.~\ref{limitplot}, 
experiments under construction should probe a large fraction of 
parameter space
allowed by minimal supersymmetric theory and experimental 
constraints~\cite{bottino,baltz01,ellis01}.
Interestingly, if the sign of $\mu$, the Higgs mixing parameter in 
the superpotential, is positive, the WIMP mass must be $< 500$~\gev,
and the WIMP-nucleon cross section must be 
$\agt10^{-9}$~pb~\cite{baltz01,ellis01}, 
possibly
making the entire parameter space accessible to a ton-scale
WIMP-detection experiment.
CryoArray would be a 1-ton deployment of semiconductor cryogenic
detectors, of a type similar to that in use in CDMS II~\cite{saabthesis}. 
This
experiment would be sensitive to a WIMP-nucleon cross section
$\sigma \approx 6\times 10^{-10}$~pb,
corresponding to a signal of a few 
WIMP interactions 
per 100~kg years.  

The detectors of CDMS or CryoArray measure phonons and
charge carriers separately for each interaction in order to
allow rejection of the otherwise 
dominant electron-recoil background events.
The background discrimination of these 
detectors has already been demonstrated to be so 
good~\cite{r19prlprd,saabthesis} that 
instrumenting a one-ton detector is fully justified.  
Below, we review 
in detail the expected performance of CryoArray, and conclude with 
a short discussion of the challenge to build such an experiment at a 
reasonable cost.

\begin{table*}[th]
\caption{
Mean single-detector event rates and counts (\#) between 15--45~keV recoil energy
in the Ge detectors of CDMS I, CDMS II, 
and CryoArray. 
Rates are listed in units of mdru ($10^{-3}$~keV$^{-1}$~kg$^{-1}$~day$^{-1}$). 
Values listed for CDMS I and preliminary rejection efficiencies 
listed for CDMS II at the Stanford Underground Facility (SUF) have 
been achieved; values listed for CDMS II at Soudan and for CryoArray
(at Soudan or at a potential deeper site)
are projections. }
\begin{tabular}{llrrlrrrrl} 
\hline
&  & & \multicolumn{1}{c}{Event} & \multicolumn{1}{c}{Exposure} &   
       \multicolumn{1}{c}{Raw}  &  &
	      \multicolumn{1}{c}{After} & \multicolumn{2}{c}{After}  \\ 
&  & \multicolumn{1}{c}{Depth} & \multicolumn{1}{c}{Rate} & 
     \multicolumn{1}{c}{(1000} & \multicolumn{1}{c}{Events}  & 
     Rejection &
       \multicolumn{1}{c}{Reject} & \multicolumn{2}{c}{Subtraction} \\ 
&  Site & \multicolumn{1}{c}{(mwe)} & \multicolumn{1}{c}{(mdru)} 
     & \multicolumn{1}{c}{kg day)} &   
       \multicolumn{1}{c}{(\#)}  & Efficiency &
       \multicolumn{1}{c}{(\#)} & \multicolumn{1}{c}{(\#)} 
       & \multicolumn{1}{c}{(mdru)} \\ \hline
\multicolumn{5}{l}{Photons}         &        &           &     &    \\ \hline
CDMS I    & SUF    &   16 & ~800   &  ~~~0.016 &    384 &   99.96\% & 
0.1 &  0~ &  ~~0 \\ 
CDMS II   & SUF    &   16 & ~800   &  ~~~0.04  &    960 &   99.97\% & 
0.3 &  0~ &  ~~0 \\ 
CDMS II   & Soudan & 2080 & ~260   &  ~~~2.50  &  19500 &   99.97\% & 
6.5 &  5~ &  ~~0.07 \\ 
CryoArray &        &      & ~~13   &  500      & 195000 &   99.97\% & 
65~~  & 15~ &  ~~0.001 \\ \hline
\multicolumn{5}{l}{Electrons}      &        &           &     &   \\ \hline
CDMS I    & SUF    &   16 & ~~300   &  ~~~0.016 &    145 &   95.00\% & 
7~~  &  7~ &  15 \\ 
CDMS II   & SUF    &   16 &  ~~80   &  ~~~0.04  &     96 &   95.00\% & 
5~~  &  5~ &  ~~4 \\ 
CDMS II   & Soudan & 2080 &  ~~20   &  ~~~2.50  &   1500 &   95.00\% & 
75~~  & 15~ &  ~~0.2 \\ 
CryoArray &        &      &  ~~~1   &  500      &  15000 &   99.50\% & 
75~~  & 15~ &  ~~0.001 \\ \hline
\multicolumn{5}{l}{Neutrons produced in shield} &     &           &     &   \\ \hline
CDMS I    & SUF    &   16 & 2200  &  ~~~0.016 &   1000 &   99.90\% & 
1~~   & 0~ &  ~~0 \\ 
CDMS II   & SUF    &   16 & 1000  &  ~~~0.04  &   1200 &   99.95\% & 
0.5 & 0~ &  ~~0 \\ 
CDMS II   & Soudan & 2080 & ~~0.5   &  ~~~2.50  &     38 &   99.90\% & 
0~~   & 0~ &  ~~0 \\ 
CryoArray & Soudan & 2080 & ~~0.5   &  500      &   7500 &   99.90\% & 
8~~   & 6~ &  ~~0.0004 \\ 
Cryoarray & NUSL   & 4500 & ~0.020 &  500      &    300 &   99.90\% & 
0~~   & 0~ &  ~~0 \\ \hline
\multicolumn{5}{l}{Neutrons produced in cavern rock} &     &             &     &   \\ \hline
CDMS I    & SUF    &   16 &  50 &  ~~~0.016 &    24 & $\sim50$\% & 
12~~ & 8~ &  17 \\ 
CDMS II   & SUF    &   16 &  22 &  ~~~0.04  &    26 & $\sim50$\% & 
13~~ & 8~ &  ~~7 \\ 
CDMS II   & Soudan & 2080 &  0.22 &  ~~~2.50  &    16 & $\sim50$\% & 
8~~ & 6~ &  ~~0.08 \\ 
CryoArray & Soudan & 2080 &  0.22 &  500      &  3300 & $\sim50$\% & 
1650~~ & 80~ &  ~~0.005 \\ 
CryoArray & NUSL   & 4500 &  0.01 &  500      &   150 & $\sim50$\% & 
75~~ & 18~ &  ~~0.001 \\ 
CryoArray & NUSL   & 7200 & 0.0004 & 500      &     6 & $\sim50$\% & 
3~~ & 3~ &  ~~0.0002 \\ 
\hline
\end{tabular}  
\label{backgrounds}
\end{table*}

\section{Expected Backgrounds}

Table~\ref{backgrounds} lists photon, electron, and
neutron background event rates in the energy range 15--45~keV
for actual exposures of  
CDMS~I~\cite{r19prlprd} and CDMS~II at Stanford~\cite{saabthesis}, 
and for  projections for CDMS~II at Soudan
and for CryoArray.   
The goals of CryoArray 
reflect a factor 20  improvement in backgrounds, and a factor 10
improvement in electron discrimination, over the goals for CDMS~II. 
The low-energy threshold is at 15~keV 
(rather than 5 or 10~keV) to avoid the contributions of the 
cosmogenic activation peaks and tritium beta spectrum in Ge near 10~keV.

The first columns of the table show the radioactivity event rates and the total
numbers of events expected during the tabulated exposures.
The other columns show how event-by-event discrimination can be used 
not only to reject most of the background, but also to subtract part
of the residual background if the detector response is well enough
understood~\cite{gaitskelldiscrim}. 
The results of this background subtraction include the  
effects of a 5\% systematic uncertainty in the 
detector discrimination. 
A muon veto provides the neutron rejection
(see Section~\ref{neutrons}). 
However, the neutron
subtraction is performed based on the observed population of
multiple-detector neutron events.

\subsection{Photon Background}
\label{photons}

As shown in Fig.~\ref{discrimination},
the original
CDMS~II and CryoArray goals~\cite{onetonne}
for rejection of photons have already been exceeded.
Based on the demonstrated rejection,
the original photon-background goal for CDMS~II 
(15 events after subtraction) 
has already been met,
even without the expected 3$\times$ reduction in rate from 
increasing the experiment's depth. 
Furthermore, the rejection efficiencies 
in Table~\ref{backgrounds} are 
statistically limited lower limits. 
and are probably overly conservative. 

Based on the demonstrated rejection, 
CryoArray's 
post-subtraction
background goal for photons is achievable
merely by reducing the raw photon rate to 13~mdru,
only $2\times$ lower than that currently achieved by IGEX~\cite{igex2002}
in HPGe detectors in the same energy range.  
This modest improvement is expected  
due to the increased detector self-shielding and 
multiple-scatter rejection 
from
increased mass and many detectors.

\subsection{Electron Background}
\label{electrons}

\begin{figure}
\begin{center}
\includegraphics [width=.45\textwidth]{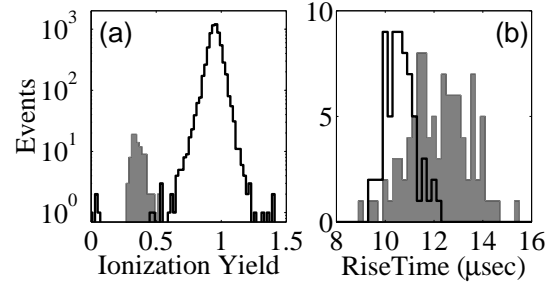}
\end{center}
\caption{Discrimination of background events in the energy range 10--100~keV
as demonstrated with CDMS~II calibration data~\cite{saabthesis}
from sources of photons (unfilled histograms) and neutrons 
(filled histograms). (a)
Discrimination of photons is
$>99.97$\% efficient, limited by statistics.
(b) Applying a cut based on phonon risetimes can remove
$>97$\% of the low-ionization-yield events (mostly electrons) while keeping 
$>50$\% of the neutron events.
 }
\label{discrimination}
\end{figure}

The beta electron background represents a greater challenge. 
The current observed background rate 
for CDMS~II
at SUF
is 80~mdru, 
$4\times$ above the CDMS~II Soudan goal
of 20~mdru. 
Some of the current beta
contamination may be due to exposure of the detectors
to a leaking calibration 
source during testing. 
Use of such sources has been restricted.
Moving to the deeper site at Soudan will remove essentially all 
cosmogenically produced betas, potentially the dominant beta background.
Furthermore,  the
discrimination performance of the new detectors (see 
Fig.~\ref{discrimination}b)
may well be sufficient even if CDMS~II 
does not reach the absolute beta background goal.

For CryoArray, self-shielding will eliminate the beta 
backgrounds from materials surrounding the detector stacks.
Nevertheless, improving the beta background by 
a further factor of 
20 will 
require sensitive screening of the detectors and nearby materials. 
While screening to the required level of 
$0.02$~counts~keV$^{-1}$~m$^{-2}$~day$^{-1}$ 
for betas with $E > 50$~keV
appears achievable~\cite{betaScreening}, 
cleaning the materials may require 
new techniques.
Improving the electron discrimination to 99.5\% appears achievable through 
improvements of the signal-to-noise of the
detectors. 

%

\subsection{Neutron Background}
\label{neutrons}

The low-energy ($<$10~MeV) neutrons from 
$(\alpha$,n) and fission processes
in the rock are trivially stopped by hydrogenous
shielding.  
Such shielding reduces the low-energy neutron flux by $\sim$10$\times$
per 10~cm of material,
so 70~cm of such shielding should make this neutron flux negligible. 

Other neutrons  
arise from muons interacting inside the experiment's
shielding or in the cavern rock.
The former can be tagged  using a local active muon veto
directly around the passive shielding.  
A muon veto efficiency of 80\% is adequate at 
Soudan (2080 mwe) to ensure that these ``shield'' 
neutrons 
contribute negligibly to the CDMS~II background.  
The demonstrated CDMS~II muon veto efficiency of 99.95\%
would even be sufficient to limit this 
neutron background to a small enough rate at Soudan for CryoArray.
A deeper site such as Gran Sasso (at $\sim$4~kmwe) would 
reduce the neutron flux by 
$\sim$25$\times$, 
making the muon-veto requirement very modest. 

The high-energy neutrons generated in the cavern rock 
by spallation processes of high-energy muons
are difficult both to veto and to stop. 
With the CDMS II polyethylene shielding sandwiching Pb shielding,
$\sim$8 secondary neutrons from these ``punch-through'' neutrons 
should scatter in single detectors
during the 2500 kg-day exposure, based on extensive Monte
Carlo studies~\cite{yellinpriv,onetonne}. 

Because this neutron flux for deep sites is approximately proportional 
to the muon flux~\cite{pdb00},
increased depth is the easiest way to reduce this neutron background.
Siting CryoArray at a depth of 4~kmwe would likely 
reduce this background sufficiently, while siting it at 7~kmwe should
essentially remove this background.
If the experiment were sited at $\alt 2000$~mwe 
(e.g. WIPP or Soudan), either a much thicker liquid-scintillator
buffer would be required around the detector in order to tag high-energy
neutrons, or
the cavern rock itself (or an outer heavy shield) 
would need to be instrumented  with
additional veto detectors in order to catch some part of the  shower
associated with the muon that generated the neutron. 

\section{
Conclusions
}

CryoArray should work well enough 
to probe WIMP-nucleon cross sections $\sigma \alt 10^{-9}$~pb
once it is built.
The most significant challenge at this point is 
determining how to build it at a reasonable cost.
For CDMS~II,
the 42 detectors are being built 
using university 
facilities and
technicians under the close supervision of CDMS physicists. 
Detectors are tested at least twice 
in batches of $\leq3$ 
in university dilution refrigerators. 
This time-consuming and physicist-intensive procedure 
is not practical for 
CryoArray's $\sim$ 500--2000
detectors. 
An industrial approach is needed. 

In order to ``out-source'' this effort, we will 
need
to engineer and specify a stable and reliable high-yield process.
Since it is not feasible to test each individual module
prior to deployment in the array, the yield must be \agt90\%.
This goal is ambitious; current detector yields are $\sim$50\%.
We expect to continue to make gains 
in understanding pathologies in the processing steps 
during the remaining CDMS~II construction period, which
is scheduled for completion at the end of 2003. 
In 1--2 years we will have substantially more experience
in fabricating and operating CDMS detectors. 
The knowledge gained
will be essential for establishing a realistic plan for building
CryoArray and assessing its likely performance. 
This knowledge will
likely point the way to additional laboratory work to be carried out
before a full-scale proposal is considered.

\section*{Acknowledgments}

We thank our CDMS collaborators for innumerable useful discussions.
This work is supported 
by the University of California, Berkeley,
under Cooperative Agreement No. AST-99-78911, and
by the National Science Foundation under Grant No. PHY-9722414. 
Some of this work was done at the Aspen Center for Physics.

\end{document}